\documentclass{article}
\usepackage{microtype}
\usepackage{dcase2022,amsmath,graphicx,url,times,booktabs, tabularx}
\usepackage{color}
\usepackage{siunitx}
\usepackage[colorlinks,linkcolor=black,urlcolor=black,citecolor=black]{hyperref}
\usepackage{soul}
\usepackage{booktabs}
\usepackage{multirow}
\usepackage{amsmath,amssymb,stmaryrd}   

\let\OLDthebibliography\thebibliography
\renewcommand\thebibliography[1]{
  \OLDthebibliography{#1}
  \setlength{\parskip}{0pt}
  \setlength{\itemsep}{1.5pt plus 0.1ex}
}
\title{Continual Learning For On-Device Environmental Sound Classification}

\name{
      Yang Xiao$^{1,*}$,
      Xubo Liu$^{2,}
      \sthanks{The first two authors contributed equally to this work.}$,
      James King$^{2}$,
      Arshdeep Singh$^{2}$,
      Eng Siong Chng$^{1}$,}
      \secondlinename{
      Mark D. Plumbley$^{2}$,
      Wenwu Wang$^{2}$
      }

\address{$^1$ School of Computer Science and Engineering, Nanyang Technological University, Singapore\\
        $^2$ Centre for Vision, Speech and Signal Processing (CVSSP), University of Surrey, UK
 }

\begin{document}

\ninept
\maketitle

\begin{sloppy}

\begin{abstract}
Continuously learning new classes without catastrophic forgetting is a challenging problem for on-device environmental sound classification given the restrictions on computation resources (e.g., model size, running memory). To address this issue, we propose a simple and efficient continual learning method. Our method selects the historical data for the training by measuring the per-sample classification uncertainty. Specifically, we measure the uncertainty by observing how the classification probability of data fluctuates against the parallel perturbations added to the classifier embedding. In this way, the computation cost can be significantly reduced compared with adding perturbation to the raw data. Experimental results on the DCASE 2019 Task 1 and ESC-50 dataset show that our proposed method outperforms baseline continual learning methods on classification accuracy and computational efficiency, indicating our method can efficiently and incrementally learn new classes without the catastrophic forgetting problem for on-device environmental sound classification.
\end{abstract}


\section{Introduction}
\label{sec:intro}
Environmental sound classification aims to categorize audio recordings into pre-defined environmental sound classes \cite{piczak2015environmental}. Recently, on-device environmental sound classification \cite{Singh2022, singh2022passive, choi2022temporal} has attracted increasing research interest, as shown in Task 1 of Detection and Classification of Acoustic Scenes and Events (DCASE) 2022 Challenge: ``Low-Complexity Acoustic Scene Classification" \cite{martin2022low}. Such a sound classification system with low computation-complexity can be deployed on mobile and embedded platform for many real-world audio applications, such as acoustic surveillance \cite{radhakrishnan2005audio}, bio-acoustic monitoring \cite{Liu2022a} and multimedia indexing \cite{kiranyaz2006generic}.

Most existing environment sound classification models \cite{piczak2015environmental, singh2022passive, choi2022temporal, salamon2017deep, sun2022deep} are trained with limited sound classes, which cannot directly adapt to new sound classes. When model developers want to expand the categories of environmental sounds to be classified, one way to do this is to fine-tune the model with new classes of data \cite{hou2019domain,hou2020multi}. However, this method may discard previously learned knowledge during the fine-tuning process: this is also known as the catastrophic forgetting problem \cite{mccloskey1989catastrophic}. Another possible solution is to re-train sound classification models with a mixture of historical and new data. However, this method is resource- and time-consuming in real-world on-device scenarios. 
As the solution based on re-training is computationally expensive, it is important to design efficient and effective methods to adapt the trained on-device sound classification model to new sound classes.

Continual learning (CL) \cite{awasthi2019continual,zhang2021serverless,huang2021modelci} aims to continuously learn new knowledge over time while retaining and reusing previously learned knowledge. Existing CL methods can be generally divided into two categories: regularization-based methods \cite{kirkpatrick2017overcoming, zenke2017continual} and replay-based methods \cite{hsu2018re, rebuffi2017icarl}. Regularization-based methods use a regularization loss to preserve previously learned model parameters when learning new knowledge. Replay-based methods use a memory update algorithm (MUA) \cite{vitter1985random, rebuffi2017icarl, bang2021rainbow} to sample a few informative examples from historical data. The selected examples are used to preserve information about old classes when training new classes. Recently, replay-based CL methods have shown promising results outperforming regularization-based methods in audio tasks such as keywords spotting \cite{huang2022progressive, xiao2022rainbow} and sound event detection \cite{wang2019continual}. However, CL in on-device applications, such as on-device environmental sound classification, has received less attention in the literature, which is the focus in this paper. The on-device scenarios are often associated with restrictions in storage and memory space \cite{singh2022passive}, which can pose challenges to replay-based CL which relied on external memory to restore historical data. As a result, the sound classification models that can be operated on the device may be limited in their capacities, thus prone to forgetting old knowledge when continuously learning new sound classes.


In this work, we investigate the replay-based CL (RCL) methods for on-device environmental sound classification. We first study the performance of existing memory update algorithm (MUA) methods such as \textit{Reservoir} \cite{vitter1985random}, \textit{Prototype} \cite{rebuffi2017icarl} and \textit{Uncertainty} \cite{bang2021rainbow} (as described in Section \ref{sec:mua}) on RCL for on-device environmental sound classification.
We empirically demonstrate that \textit{Uncertainty} \cite{bang2021rainbow} method performs best in our scenario. Furthermore, we propose \textit{Uncertainty++}, a simple yet and efficient MUA method based on \textit{Uncertainty} method. Different the \textit{Uncertainty} method, our proposed \textit{Uncertainty++} introduces the perturbations to the embedding layer of the classifier. As a result, the computation cost (e.g., running memory and time) can be significantly reduced when measuring the data uncertainty. We evaluate the performance of our method on the DCASE 2019 Task1 \cite{tau2019} and the ESC-50 \cite{piczak2015esc} datasets with on-device model BC-ResNet-Mod ($\sim$86k parameters) \cite{kim2021broadcasted,Kim2021b}. Experimental results show that \textit{uncertainty++} outperforms the existing MUA methods on classification accuracy, indicating its potential in real-world on-device audio applications. Our proposed method is model-independent and simple to apply. Our code is made available at the GitHub\footnote{\url{https://github.com/swagshaw/ASC-CL}}.


%

The remainder of this paper is organized as follows. Section \ref{sec:method} introduces the continual learning method we proposed for on-device environmental sound classification. Section \ref{sec:exp} presents the experimental settings and the evaluation results. Conclusions and future directions are given in Section \ref{sec:conclusion}.

\section{Method}
\label{sec:method}

This section first describes replay-based continual learning and four memory update algorithms, and then introduces the proposed \textit{uncertainty++} algorithm.

\subsection{Replay-based continual learning}
\label{sec:mua}
Follow the continual learning setting \cite{zenke2017continual,wang2019continual,awasthi2019continual} of environmental sound classification, we assume that the model M should identify all classes in a series of tasks $T=\{\tau_0,\dots,\tau_t\}$ without catastrophic forgetting. For each task $\tau\in T$, we have input pairs $(x,y)$ and classes $C=\{c|c\in\tau_y\}$, where $x$ denotes audio waveforms and $y$ are classes $c\in \tau_C$. We aim to minimize a cross-entropy loss of all classes $C$ present in the current task $\tau$ formulated as:
\begin{equation}
L_{CE}(\tau) = \sum\limits_{c\in \tau_C} \tau_{y_c} log \frac{exp{(M(\tau_x)}_c)}{\sum\limits_{c\in C} exp{(M(\tau_x)}_c)},
\label{eq1}
\end{equation}
Where $M(\tau_x)$ denotes the output of $\tau_x$ on out model $M$.

The parameters learned from the previous task are potentially overwritten after learning the new class, also known as catastrophic forgetting. To mitigate this issue, we introduce replay-based methods. The replay-based methods utilize a region of the memory which is called `replay buffer' to temporarily store the historical training samples to maintain the performance. 

Re-training sound classification models with the mixture of the whole historical and new data is resource- and time-consuming in real-world on-device scenarios. To mitigate this issue, the replay-based methods access only a subset of the historical data to save the storage space. In this case, how to select the part of samples to the replay buffer by the memory update algorithm is the key.

Specifically, in the training of task $\tau_t$, the replay buffer stores the selected training samples from the previous $t-1$ learned task(s) $\{\tau_0,\tau_1,\dots,\tau_{t-1}\}$, and builds the training data buffer $\hat D_t$ for task $\tau_t$ formulated as:

\begin{equation}
\label{eq2}
\hat D_t = g(\hat D_{t-1})\cup D_t,
\end{equation}
where $g$ is the memory update algorithm \cite{xiao2022rainbow}, $\hat D_{t-1}$ is the training data buffer for task $\tau_{t-1}$, and $D_t$ is the incoming data for the new task. 

\subsubsection{Memory update algorithm (MUA)}
We introduce four memory update algorithms in the literature. Generally, we assume that the memory update should select $L$ samples from the training data $\hat D_{t-1}$ of the previous task $\tau_{t-1}$ for the training of the task $\tau_t$. 

\noindent\textbf{\textit{Random}} \cite{hsu2018nr} memory update algorithm selects $L$ new samples $\{(x_1,y_1),(x_2,y_2),\dots,(x_L,y_L)\}$ for the next task randomly from the candidates $\hat D_{t-1}$ into replay buffer.

\noindent\textbf{\textit{Reservoir}} \cite{vitter1985random} memory update algorithm conducts uniform sampling from $\hat D_{t-1}$. Specifically, the reservoir algorithm initializes the replay buffer indexed from 1 to $L$, containing the first $L$ items $\{(x_1,y_1),(x_2,y_2),\dots,(x_L,y_L)\}$ of the candidates. When updating replay buffer from the candidates, for each sample, the reservoir algorithm generates a random number $m$ uniformly in $\{1,\dots, len(\hat D_{t-1})\}$. If $m \in \{1,\dots,L\}$, then the sample with the index $m$ in the replay buffer is replaced with the sample $\hat D_{t-1}[m]$.

\noindent\textbf{\textit{Prototype}} \cite{rebuffi2017icarl} 
 memory update algorithm selects the samples from $\hat D_{t-1}$ where the embedding of the classifier is close to the embedding mean of its own class. Specifically, the algorithm first groups the $\hat D_{t-1}$ into subsets as $D_c, c=1\dots N^t$ by unique classes, where $N^t$ denotes the total numbers of unique classes in the $\hat D_{t-1}$ set. Then the algorithm uses the current model to extract the embedding of the candidates for each $D_c$ and calculates the class mean by the embedding as the average feature vector. For each class, the algorithm selects the samples of the candidates so that the average feature vector over the replay buffer provides best approximate to the average feature vector over all the samples of the corresponding class.

\noindent\textbf{\textit{Uncertainty}} \cite{bang2021rainbow} memory update algorithm selects the sample by the uncertainty of the sample through the inference by the classification model. Specifically, the first step groups the $\hat D_{t-1}$ in the same way as the \textit{prototype} algorithm introduced above. The second step estimates the uncertainty of each sample $x$ in $D_c$. Predictive likelihood captures how well a model fits the data, with larger values indicating better model fit. Uncertainty score can be determined from predictive likelihood \cite{gal2016dropout}. Following the derivation from \cite{gal2016dropout}, the predictive likelihood of a sample given by the model can be approximated by the Monte-Carlo (MC) integration \cite{kloek1978bayesian} method with the model outputs of perturbed samples \cite{xiao2022rainbow}, which is defined as follows:
\begin{equation}
\label{eq3}
    P(y = c \mid x) = \int p(y=c \mid \hat{x})p(\hat{x}\mid x) d\hat{x},
\end{equation}
where $x$, $\hat{x}$, $y$ denote an audio utterance of one class, the perturbed samples of $x$, and the label of $x$. Therefore, the uncertainty of the audio utterance $x$ is formulated as $u(x)$: 
\begin{equation}
\label{ux}
    u(x) \approx 1- \frac{1}{K} \sum\limits^K_{k=1} P(y = c\mid \hat{x}_k),
\end{equation}
where $K$ presents the number of the perturbations generated by perturb methods such as Audio Shift \cite{ko2015audio}, Audio PitchShift \cite{ko2015audio} and Audio Colored Noise \cite{audiomentation,kamath2002multi}. A larger $u(x)$ indicates a smaller confidence of the model in predicting the perturbed samples.
\begin{figure}[]
  \centering
  \includegraphics[width=0.9\linewidth]{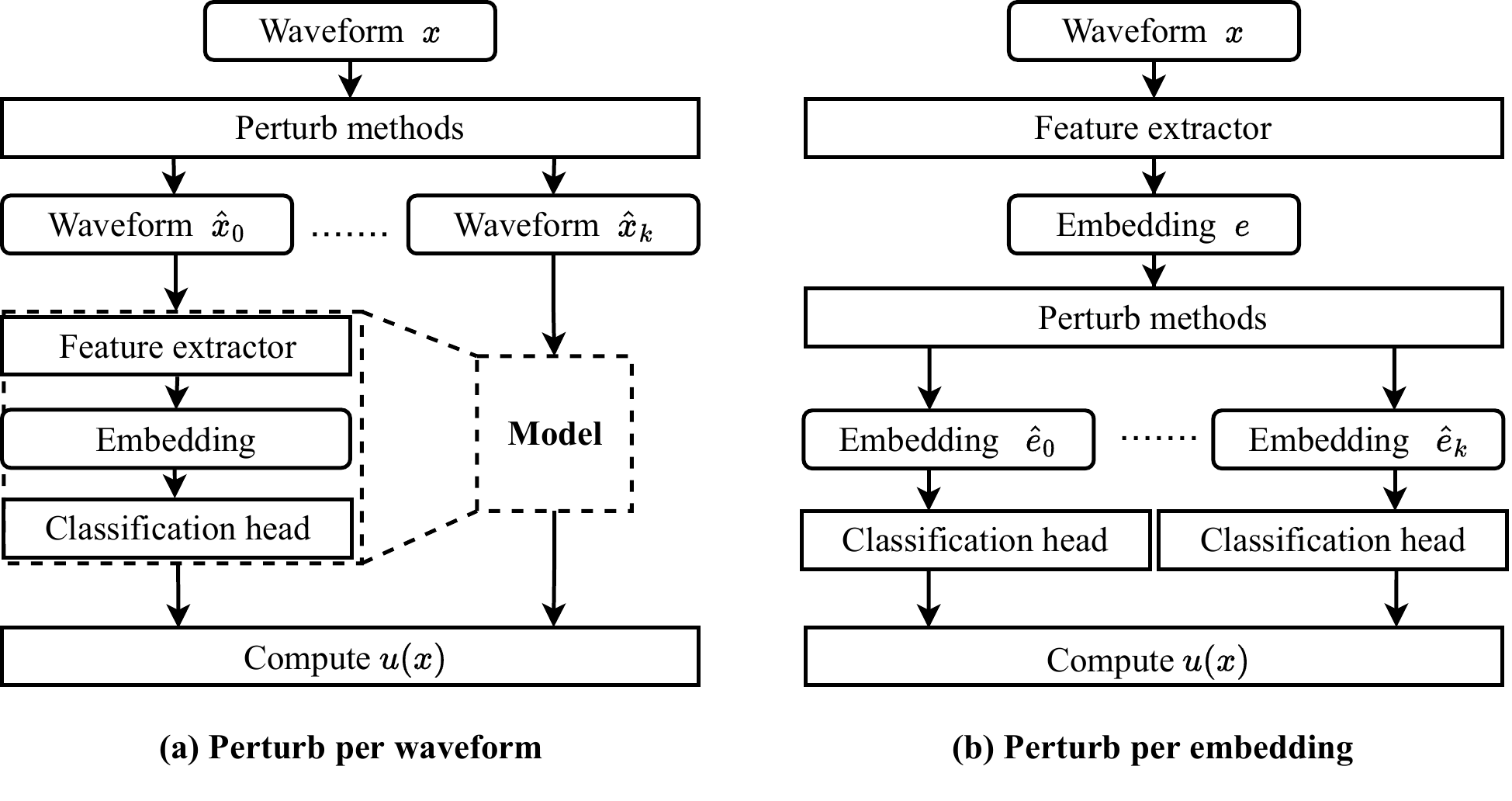}
     \vspace{-1.25em}
  \caption{\textit{Block diagram of the native uncertainty approach and our proposed approach. Specifically, the naive approach adds perturbations to $x$ by waveform and generates multiple waveform as $\hat x$. Our approach inputs the embedding $e$ and generates perturbed embedding $\hat e$ which means we only save the embedding. The output of the backbone of the model is calculated only once.  ``Compute $u(x)$" is to compute $u(x)$ by Eq. (\ref{eq3}). The $K$ refers to the number of the perturbations generated by perturb methods.}}
  \label{fig:workflow}
    \vspace{-1.25em}
\end{figure}
The third step selects $L$ examples from $D_c$ through descending the uncertainty $u(x)$ with the step size of $len(D_c)*C/L$, where $L$ is the size of the replay buffer.

Previous research \cite{xiao2022rainbow} demonstrated that the uncertainty memory update algorithm performs better than the other three algorithms on speech tasks such as keyword spotting. However, the computation cost of \textit{Uncertainty} increases linearly with the number of perturbation operations.

\subsection{Proposed MUA method (\textit{Uncertainty++)}}
 As illustrated in Figure \ref{fig:workflow}, the native uncertainty memory update algorithm requires to employ perturbation methods offline for the waveform of each sample to generate the perturbed samples first. In our proposed method, noisy perturbations are added to the pre-classifier embedding of the sample, and not to the waveform, so the output of the backbone of the model is calculated only once. Specifically, we propose a vector-wise perturbation method that adds noise with different intensities according to the variance of classifier's embedding. We denote the perturbed version of the classifier's embedding $e$ as $\hat e$, which is computed as follows:

\begin{equation}
\hat e = e + U(-\frac{\lambda}{2},\frac{\lambda}{2})*std(e),
\end{equation}
where $std(\cdot)$ stands for standard deviation, the function $U(a,b)$ represents the noise distributed uniformly from $a$ to $b$, $U(a,b)$ is a vector with the same shape as $e$, and $\lambda$ is a hyperparameter that controls the relative noise intensity. 

By the vector-wise perturbation method, we generate the perturbed embedding $\hat e$ of the embedding $e$. Finally, we input $\hat e$ to the final classification layer of the model and output $P(y=c\mid\hat{e})$ which is used to compute the uncertainty as in Eq. (\ref{eq3}). After the uncertainty is estimated, we select examples for replay as native approach. This method saves time by calculating the output of the backbone of the model only once. We also save the memory usage by replacing the perturbed raw data with the classifier's embedding which is of much smaller size as compared with the raw data.

\begin{table}[h]
\centering
\resizebox{0.45\textwidth}{!}{
\begin{tabular}{@{}ccccc@{}}
\toprule
\multirow{2}{*}{\textbf{Method}} & \multicolumn{2}{c}{\textbf{DCASE 2019 Task 1}} & \multicolumn{2}{c}{\textbf{ESC-50}} \\ \cmidrule(l){2-5} 
                        & \textbf{ACC $\uparrow$}              & \textbf{BWT} $\uparrow$             & \textbf{ACC} $\uparrow$          & \textbf{BWT} $\uparrow$         \\ \cmidrule{1-5}
\textit{Finetune}             & 0.205          & -0.276          & 0.181          & -0.307          \\ \hline
\textit{Random}               & 0.473          & -0.115          & 0.225          & -0.231          \\ \hline
\textit{Reservoir}            & 0.568          & -0.096          & 0.430          & -0.121          \\ \hline
\textit{Prototype}            & 0.559          & -0.089          & 0.482          & \textbf{-0.104} \\ \hline
\textit{Uncertainty} & 0.578 & \textbf{-0.079} & 0.477 & -0.111          \\  \hline
\textit{Uncertainty++} & \textbf{0.581} & \textbf{-0.079} & \textbf{0.500} & -0.121          \\ 
\hline
\end{tabular}}%
\caption{\textit{Accuracy (ACC) and Backward Transfer (BWT) in a comparative study of the proposed memory update algorithm.}}
  \vspace{-0.5em}
\end{table}

\begin{table}[h]
\centering
\resizebox{0.45\textwidth}{!}{
\begin{tabular}{@{}lccccc@{}}
\toprule
\multicolumn{1}{c}{\multirow{2}{*}{\textbf{Method}}}  &
\multicolumn{1}{c}{\multirow{2}{*}{\textbf{K}}} &
\multicolumn{2}{c}{\textbf{DCASE 2019 Task 1}} & \multicolumn{2}{c}{\textbf{ESC-50}} \\ \cmidrule(l){3-6} 
                     &   & \textbf{ACC $\uparrow$}              & \textbf{BWT} $\uparrow$             & \textbf{ACC} $\uparrow$          & \textbf{BWT} $\uparrow$         \\ \cmidrule{1-6}
\multicolumn{1}{c}{} & 2                 & 0.557                 & -0.101                & 0.461
             & \textbf{-0.111}
            \\
\multicolumn{1}{c}{\textit{Uncertainty-Shift}} & 4                 & 0.575                 & -0.103                & 0.476
             & -0.118
             \\
\multicolumn{1}{c}{} & 6                 & 0.567                 & -0.079                & 0.477
             & -0.118
            \\ \hline
\multicolumn{1}{c}{} & 2                 & 0.560                 & -0.100                & 0.465
             & -0.118            \\
\multicolumn{1}{c}{\textit{Uncertainty-Noise}} & 4                 & 0.535                 & -0.104                & 0.473
             & -0.118
            \\
\multicolumn{1}{c}{} & 6                 & 0.578                 & -0.079                & 0.458
             & -0.120
            \\ \hline
\multicolumn{1}{c}{} & 2             & 0.571                 & -0.102                & \textbf{0.500}
              & -0.121
            \\
\multicolumn{1}{c}{\textit{Uncertainty++}} & 4            & 0.548                 & -0.103                & 0.481
             & -0.114
            \\
\multicolumn{1}{c}{} & 6             & \textbf{0.581}                 & \textbf{-0.079}                 & 0.484
              & -0.119            \\ \bottomrule
\end{tabular}}
\caption{\textit{Accuracy (ACC) and Backward Transfer (BWT) in a comparative study of the proposed perturbation method. The $K$ refers to the number of the perturbations generated by perturbation methods.}}
\end{table}

\section{Experiments}
\label{sec:exp}

\subsection{Environmental sound classification model}
For the on-device environmental sound classification model, we use BC-ResNet-Mod \cite{Kim2021b} which is an adaptation of the BC-ResNet \cite{kim2021broadcasted} that achieves improved results on acoustic scene classification. The BC-ResNet paradigm works via repeatedly extracting spectral and then temporal features in series. Because these spectral features are of a lower dimension than the input, this model has fewer parameters than the one by dealing with the input directly. Feature extraction is channel-wise too, and both parameter reductions have negligible impact on performance \cite{kim2021broadcasted}. For our experiments, we use BC-ResNet-Mod-4, which increases the input channel dimension to 80 before extracting spectral and temporal features.
\subsection{Datasets}
\textbf{ESC-50} consists of \num{2000} five-second environmental audio recordings \cite{piczak2015esc}. Data are balanced between \num{50} classes, with \num{40} examples per class, covering animal sounds, natural soundscapes, human sounds (non-speech), and ambient noises. The dataset has been prearranged into five folds for cross-validation.

\noindent \textbf{DCASE 2019 Task 1} is an acoustic scene classification task, with a development set \cite{tau2019} consisting of \num{10}-second audio segments from \num{10} acoustic scenes: airport, indoor shopping mall, metro station, pedestrian street, public square, the street with a medium level of traffic, traveling by tram, traveling by bus, traveling by an underground metro and urban park. In the development set, there are \num{9185} and \num{4185} audio clips for training and validation, respectively.

\subsection{Experimental setup}
\textbf{Task setting} To evaluate the performance of the proposed approach, we split the data into five tasks. Each task includes \num{2} new unique classes in DCASE 19 Task 1 and \num{10} new unique classes in ESC-50, which is unseen in previous tasks. To simulate the condition of edge devices, we set the max amount of examples as \num{500}, \num{100} samples in DCASE 19 Task 1 and ESC-50 due to the memory limitation. 

\noindent \textbf{Implementation details} The original audio clip is converted to \num{64}-dimensional log Mel-spectrogram by using the short-time Fourier transform with a frame size of \num{1024} samples, a hop size of \num{320} samples, and a Hanning window. The classification network is optimized by the Adam \cite{kingma2014adam} algorithm with the learning rate \num{1e-3}. The batch size is set to \num{32} and the number of epochs is \num{50}. 


\subsection{Evaluation metrics}
We report performances in terms of the accuracy and forgetting metric. Specifically, the \textit{Accuracy} (ACC) reports an accuracy averaged on learned classes after the entire training ends. The \textit{Backward Transfer} (BWT) \cite{lopez2017gradient} evaluates accuracy changes on all previous tasks after learning a new task, indicating the forgetting degree.  For measuring BWT, we first construct the matrix $R \in \mathbb{R}^{T\times T}$, where $R_{i,j}$ is the test classification accuracy of the model on task $\tau_j$ after observing the last sample from task $\tau_i$. After the model finished learning about each task $\tau_i$, we evaluate its BWT on all $T$ tasks, which is formulated as:
\begin{equation}
\begin{aligned}
BWT = \frac{1}{T-1} \sum\limits^{T-1}_{i=1} R_{T,i} - R_{i,i}.
\end{aligned}
  \vspace{-0.75em}
\end{equation}
There exists negative BWT when learning about some task decreases the performance on some preceding task. The smaller value of BWT indicates the higher extent of catastrophic forgetting. 
\subsection{Reference baselines}
We built five baselines for comparisons. The \textit{Finetune} training strategy adapts the BC-ResNet-Mod model for each new task without any continual learning strategies, as the lower-bound baseline. The four prior memory update algorithms of replay-based continual learning (i.e., \textit{Random}, \textit{Reservoir}, \textit{Prototype}, \textit{Uncertainty}) are introduced in Section \ref{sec:mua}. Specifically, at the perturbation stage of the uncertainty, we use two perturbation methods, namely, `\textit{uncertainty-shift}', which includes Audio Shift and Audio PitchShift, and  `\textit{uncertainty-noise}' which refers to the Audio Colored Noise perturbation method.
\section{Results}
\subsection{Experiments on MUA methods}
Table 1 presents the results on DCASE 2019 Task 1 and ESC-50 test set in terms of ACC and BWT. We compare the proposed \textit{Uncertainty++} MUA method with five baselines. We observe that the \textit{uncertainty} MUA method achieves better performance than the five baselines. Comparing with the best baseline \textit{uncertainty}, we observe that the proposed \textit{uncertainty++} method obtains 58.1\% on classification accuracy which outperforms the existing MUA methods. In addition, we observe that the \textit{Finetune} method achieves the worst ACC and BWT performance compared with other baselines, which indicates the issue of catastrophic forgetting.

We further analyze and summarize the performances of the proposed \textit{uncertainty++} method compared with the \textit{uncertainty} MUA method with different numbers of the perturbation methods in terms of ACC and BWT as shown in Table 2. The $K$ refers to the number of the perturbations generated by perturb methods. Even with only two perturbation methods, our proposed method still outperforms other two baselines. We also observe that our method under two perturbations obtains the best performance on the ESC-50 test set. Such performance might be due to the small size of the ESC-50.


\subsection{Comparative experiments on computation time for \textit{Uncertainty} and \textit{Uncertainty++}}
We further report the Average Time for the proposed method when there is an increasing number of perturbations.
The Average Time measures the additional time brought by the memory update algorithms in each task. As shown in Table 3, even with $6$ perturbations, the Average Time of the \textit{uncertainty++} is still less than \num{60}s. This can be explained by the fact that the our proposed method can limit the growth of the additional training time. We also observe that our proposed method outperforms other baselines in any number of perturbations, which indicates our proposed method is computationally more efficient. In addition, the average time of \textit{uncertainty-shift} is much longer than others. Because the Audio Shift and Audio PitchShift perturbations takes more time than simply adding noise.



\begin{table}[]
\centering
\begin{tabular}{@{}lccc}
\toprule
\multicolumn{1}{c}{\textbf{Method}}  & \multicolumn{1}{c}{\textbf{K}} & \multicolumn{1}{c}{\textbf{Average Time (s) $\downarrow$}}         \\ \cmidrule{1-3}
\multicolumn{1}{c}{} & 2                 & 1221.7                
            \\
\multicolumn{1}{c}{\textit{Uncertainty-Shift}} & 4                 & 2205.1                
             \\
\multicolumn{1}{c}{} & 6                 & 2926.1                
            \\ \hline
\multicolumn{1}{c}{} & 2                 & 246.2                           \\
\multicolumn{1}{c}{\textit{Uncertainty-Noise}} & 4                 & 390.8               
            \\
\multicolumn{1}{c}{} & 6                 & 506.3                
            \\ \hline
\multicolumn{1}{c}{} & 2 & 44.0            
            \\
\multicolumn{1}{c}{\textit{Uncertainty++}} & 4            & 48.5               
            \\
\multicolumn{1}{c}{} & 6             & 55.1                \\ \bottomrule
\end{tabular}
\caption{\textit{Average Time (s) in a comparative study of the proposed \textit{uncertainty++} method. The $K$ refers to the number of the perturbations generated by perturbation methods.}}
  \vspace{-1.25em}
\label{tab:time}
\end{table}

\section{Conclusions}
In this work, we have presented \textit{uncertainty++}, an efficient replay-based continual learning method for on-device environmental sound classification. Our method selects the historical data for the training by measuring the per-sample classification uncertainty on the embedding layer of the classifier. Experimental results on the DCASE 2019 Task 1 and ESC-50 datasets show that our proposed method outperforms the baseline continual learning methods on classification accuracy and computational efficiency. In future work, we plan to apply and adapt our approach to other on-device audio classification tasks such as audio tagging and sound event detection.
\label{sec:conclusion}


\section{ACKNOWLEDGMENT}
\label{sec:ack}

This work is partly supported by UK Engineering and Physical Sciences Research Council (EPSRC) Grant EP/T019751/1 ``AI for Sound'', a Newton Institutional Links Award from the British Council, titled ``Automated Captioning of Image and Audio for Visually and Hearing Impaired'' (Grant number 623805725), a Research Scholarship from the China Scholarship Council (CSC), and a PhD studentship from the Engineering and Physical Sciences Research Council (EPSRC) Doctoral Training Partnership EP/T518050/1. For the purpose of open access, the authors have applied a Creative Commons Attribution (CC BY) licence to any Author Accepted Manuscript version arising.


\bibliographystyle{IEEEtran}
\bibliography{template.bbl}

%
%
%
%
%
%
%
%
%

\end{sloppy}
\end{document}